\documentclass[prd,preprint,superscriptaddress,nofootinbib,longbibliography]{revtex4-1}
\usepackage{graphicx}
\usepackage{amsmath}
\usepackage{amsfonts}
\usepackage{mathtools}
\usepackage{color}
\usepackage{amssymb}
\usepackage{dcolumn}
\usepackage{bm}
\usepackage{braket}
\usepackage{microtype} 
\usepackage[linktoc=all]{hyperref}
\usepackage{cleveref}
\usepackage{yfonts}

\numberwithin{equation}{section}

\newcommand{\be}{\begin{equation}}
\newcommand{\ee}{\end{equation}}
\newcommand{\bea}{\begin{eqnarray}}
\newcommand{\eea}{\end{eqnarray}}
\newcommand{\beas}{\begin{eqnarray*}}
\newcommand{\eeas}{\end{eqnarray*}}

\def\({\left(}
\def\){\right)}

\makeatletter
\DeclareRobustCommand{\rcite}[1]{%
\rcite@aux#1,\@nil{#1}%
}
\usepackage{amsmath}
\usepackage{amsfonts}
\usepackage{xcolor}
\usepackage{bbold}
\usepackage{microtype}
\usepackage{hyperref}
\usepackage{cleveref}
\newcommand{\dd}{\mathrm{d}}
\newcommand{\Mp}{M_\mathrm{Pl}}
\newcommand{\sdg}{\sqrt{-g}}

\newcommand{\gs}{g_s}
\newcommand{\mb}{\mathcal{B}}

\makeatletter
\DeclareRobustCommand{\rcite}[1]{%
  \rcite@aux#1,\@nil{#1}%
}
\def\rcite@aux#1,#2\@nil#3{%
  \if\relax#2\relax
    Ref.~\cite{#3}%
  \else
    Refs.~\cite{#3}%
  \fi
}
\makeatother

\begin{document}

\title{Non-canonical kinetic structures in the swampland}
\author{Adam R. Solomon}
\email{adamsolo@andrew.cmu.edu}
\affiliation{Department of Physics \& McWilliams Center for Cosmology,\\Carnegie Mellon University, Pittsburgh, Pennsylvania 15213, USA}
\author{Mark Trodden}
\email{trodden@physics.upenn.edu}
\affiliation{Center for Particle Cosmology, Department of Physics and Astronomy, University of Pennsylvania, 209 S. 33rd St., Philadelphia, PA 19104, USA}
\date{\today}

\begin{abstract}
	\noindent 
We consider how the swampland criteria might be applied to models in which scalar fields have nontrivial kinetic terms, particularly in the context of $P(\phi,X)$ theories, popularly used in approaches to inflation, to its alternatives, and to the problem of late-time cosmic acceleration. By embedding such theories in canonical multi-field models, from which the original theory emerges as a low-energy effective field theory, we derive swampland constraints, and study the circumstances under which these might be evaded while preserving cosmologically interesting phenomenology. We further demonstrate how these successes are tied to the phenomenon of {\it turning} in field space in the multi-field picture. We study both the general problem and specific examples of particular interest, such as DBI inflation. 
\end{abstract}

\maketitle


\section{Introduction}

Cosmological model-building typically works, explicitly or otherwise, with ``bottom-up" effective field theories (EFTs).\footnote{See, e.g., \rcite{Cheung:2007st,Weinberg:2008hq,Carrasco:2012cv,Solomon:2017nlh}.} According to the standard lore, one constructs a bottom-up EFT by deciding on some particle content and some set of symmetries, and then writing down the most general action consistent with those symmetries, organized in an appropriate expansion. This approach has a number of virtues: it allows one to work with low-energy theories in a good deal of generality, and it typically requires only a finite number of operators in order to compare to observations at a given level of precision. Nevertheless, the number of parameters an EFT has can be greater than can be constrained with a given set of observations, and it is interesting and useful to ask whether these parameters can be further restricted by theoretical considerations.

A variety of programs have been developed to constrain the space of low-energy effective theories using guidance from high energies. The general approach is to explore whether there are effective field theories that appear consistent, but nevertheless do not possess an ultraviolet (UV) completion.\footnote{Of course, this is a very general question, and caveats exist, e.g., the UV completion one has in mind may be required to satisfy certain properties, such as being Lorentz-invariant, or arising from a string theory.} These programs include the weak gravity conjecture \cite{ArkaniHamed:2006dz}, positivity bounds \cite{Adams:2006sv}, and the swampland \cite{Vafa:2005ui}.

In recent years, string theoretic arguments have been used to suggest that the string landscape may exclude a large ``swampland" of low-energy effective field theories which, remarkably, would include most models with quasi-de Sitter vacua, which are typically used to obtain inflation, or to model dark energy \cite{Obied:2018sgi,Agrawal:2018own}. Many such conjectures have been proposed, with two of the most prominent being {\it the de Sitter conjecture} and {\it the distance conjecture}, formulated as follows. 

Consider a low-energy effective theory consisting of a set of scalar fields $\Phi^a$ with a field-space metric $\mathcal{G}_{ab}(\Phi)$ and a potential $V(\Phi)$,
\begin{equation}
S = \int\dd^4x\sdg\left[\frac{\Mp^2}{2}R-\frac12\mathcal{G}_{ab}(\Phi)\partial_\mu\Phi^a\partial^\mu\Phi^b-V(\Phi)\right].  \label{eq:sigma-action}
\end{equation}
The de Sitter conjecture proposes that the gradient of the potential $V$ is bounded as
\begin{equation}
|\nabla V| \equiv \sqrt{\mathcal{G}^{ab}\partial_aV\partial_bV} \geq \frac c \Mp V \label{eq:dSconj}
\end{equation}
for some $\mathcal{O}(1)$ constant $c$, and with $\partial_a\equiv\partial/\partial\Phi^a$. The distance conjecture states that the field excursion is sub-Planckian,
\begin{equation}
|\Delta\Phi|\equiv\sqrt{\mathcal{G}_{ab}\Delta\Phi^a\Delta\Phi^b}<\mathcal{O}(\Mp). \label{eq:distconj}
\end{equation}
The combination of these conjectures has been argued to severely constrain single-field models of inflation and dark energy \cite{Agrawal:2018own,Akrami:2018ylq,Heisenberg:2018yae,Hertzberg:2018suv}, although multi-field models can evade these bounds through field-space effects \cite{Achucarro:2018vey}.
Refinements of these conjectures exist; for example demanding that
\begin{equation}
\operatorname{min}(\partial_a\partial_bV)\leq-\frac{c'}{\Mp^2}V
\end{equation}
for some constant $c'\sim\mathcal{O}(1)$ \cite{Garg:2018reu,Ooguri:2018wrx}.

A large class of theories of interest in cosmology are not of the $\sigma$-model form \eqref{eq:sigma-action}, but instead involve non-canonical structures. Examples include $P(\phi,X)$ theories \cite{ArmendarizPicon:1999rj,ArmendarizPicon:2000ah,ArmendarizPicon:2000dh}, galileons \cite{Nicolis:2008in,Hinterbichler:2010xn}, Horndeski models \cite{Horndeski:1974wa,Deffayet:2011gz,Kobayashi:2011nu}, and DHOST theories \cite{Langlois:2015cwa,BenAchour:2016fzp}. It is therefore natural to ask whether the swampland criteria can be applied to inflationary or dark energy models with such non-canonical kinetic terms. In this paper we demonstrate a method of applying these criteria to $P(\phi,X)$ theories, in which the action has a non-trivial, algebraic dependence on the scalar-field kinetic term $X\equiv-\frac12(\partial\phi)^2$,
\begin{equation}
S = \int\dd^4x\sdg\left(\frac{\Mp^2}{2}R+ P(\phi,X)\right). \label{eq:action}
\end{equation}
For generality, and to make contact with much of the literature, we allow arbitrary dependence on both $X$ and $\phi$. We can rewrite a $P(\phi,X)$ theory in a $\sigma$-model form by introducing an auxiliary field $\chi$ \cite{Tolley:2009fg,Elder:2014fea},
\begin{equation}
S = \int\dd^4x\sdg\left[\frac{\Mp^2}{2}R+P(\phi,\chi) + P_\chi\left(X-\chi\right)\right].
\end{equation}
Since the $\chi$ equation of motion sets $\chi=X$,\footnote{This requires $P_{\chi\chi}\neq0$, which is not a particularly restrictive condition, but one should check that it is not violated along the solution under consideration.} this action is dynamically equivalent to \cref{eq:action}. We can therefore write a ``UV extension" of the $P(\phi,X)$ theory by adding a small kinetic term for $\chi$,
\begin{equation}
S = \int\dd^4x\sdg\left[\frac{\Mp^2}{2}R+P(\phi,\chi) - \frac{1}{2\Lambda^6}(\partial\chi)^2 + P_\chi\left(X-\chi\right)\right], \label{eq:action-2field}
\end{equation}
where $\Lambda$ is a mass scale, introduced on dimensional grounds, which acts as a cutoff for the UV extension. At sufficiently low energies we can ignore the dynamics of $\chi$ and again obtain the $P(\phi,X)$ theory. The action \eqref{eq:action-2field} is of the form \eqref{eq:sigma-action}, with $\Phi^a=(\phi,\chi)$ and
\begin{align}
\mathcal{G}_{ab} &= \operatorname{diag}\left(P_\chi,\frac{1}{\Lambda^6}\right), \\
V(\Phi) &= -P+\chi P_\chi,
\end{align}
and is therefore amenable to a swampland analysis.\footnote{See \rcite{Mizuno:2019pcm} for further discussion of this idea and of other approaches to applying the swampland conjectures to $P(\phi,X)$ theories. Also see \rcite{Dasgupta:2018rtp,Mizuno:2019pcm} for previous discussions of auxiliary field techniques in the swampland context. For other approaches to applying the swampland bounds to $P(\phi,X)$ theories and their generalizations, see, e.g., \rcite{Seo:2018abc,Heisenberg:2019qxz,Lin:2018edm,Lin:2019pmj}. See \rcite{Babichev:2018twg} for an alternative auxiliary-field forumulation of shift-symmetric $P(\phi,X)=P(X)$ theories.}

The general idea here is to perform a swampland analysis for the theory~\eqref{eq:action-2field}, and then to consider the constraints on the field $\phi$ (which we might imagine as the inflaton or dark energy field) as we approach the decoupling limit of $\chi$. We will focus on applying the relevant constraints along solutions of cosmological interest, such as those describing inflation, although the general technique can be applied more broadly.

We start in \cref{sec:swamp} by formulating the swampland criterion \eqref{eq:dSconj} for general $P(\phi,X)$ theories and discussing the conditions on $P(\phi,X)$ under which it is satisfied. In \cref{sec:DBI} we apply these general considerations to DBI inflation, a prominent class of inflationary models of the $P(\phi,X)$ type. Finally we provide a physical interpretation of these results from the two-field perspective in terms of field-space turning in \cref{sec:turning}, before concluding in \cref{sec:conc}.

\section{Swampland estimates}
\label{sec:swamp}

\subsection{Swampland criterion for $P(\phi,X)$}

Our aim is to apply the de Sitter swampland criterion \eqref{eq:dSconj} to $P(\phi,X)$ theories in the UV-extended form \eqref{eq:action-2field}. We start by considering the as-yet-unfixed scale $\Lambda$, which must be chosen sufficiently large that the $P(\phi,\chi)$ theory is well-approximated by the original $P(\phi,X)$ theory on the low-energy solutions of interest. The original theory arises in the limit of infinite $\Lambda$. At first glance it appears as if the de Sitter conjecture, which depends on $\Lambda$ through the inverse field-space metric $\mathcal{G}^{ab}=\operatorname{diag}(1/P_\chi,\Lambda^6)$, is trivially satisfied in this limit. However, we are not allowed to increase $\Lambda$ arbitrarily, as we will run into strong coupling. Moreover, the distance conjecture \eqref{eq:distconj} has the opposite scaling with $\Lambda$, as it depends on $\mathcal{G}_{ab}$ rather than its inverse $\mathcal{G}^{ab}$.

To avoid such strong coupling problems and to be maximally conservative in applying the swampland conjecture, we choose to take $\Lambda$ as small as possible consistent with inflation remaining as a valid solution of the low-energy EFT. Consider the $\chi$ equation of motion,
\begin{equation}
\frac{1}{\Lambda^6}\Box\chi + P_{\chi\chi}\left(X-\chi\right) = 0,
\end{equation}
and expand $\chi$ in powers of the small parameter $1/\Lambda^6$,\footnote{The requirement is that some characteristic energy scale be less than order $\Lambda$; strictly speaking this corresponds to some small dimensionless parameter.}
\begin{equation}
\chi = X + \frac{1}{\Lambda^6}\chi_1+\mathcal{O}\left(\frac{1}{\Lambda^{12}}\right).
\end{equation}
The $\mathcal{O}(\Lambda^0)$ piece of the $\chi$ equation of motion is satisfied by our choice of the leading-order term $\chi = X+\cdots$, and the $\mathcal{O}(\Lambda^{-6})$ piece is
\begin{equation}
\Box X -P_{XX}\,\chi_1=0\Longrightarrow\chi_1=\frac{\Box X}{P_{XX}}.
\end{equation}
By insisting that this not dominate the solution for $\chi$, i.e., $\chi_1/\Lambda^6\ll X$, we obtain a lower bound on $\Lambda$,
\begin{equation}
\Lambda^6\gtrsim\frac{\Box X}{X P_{XX}}. \label{eq:minLambda}
\end{equation}
Note that the right-hand side of this expression can be evaluated entirely on the solution to the low-energy $P(\phi,X)$ theory, since by construction we have $\chi\approx X$ to this order. We will take $\Lambda$ to have the smallest value allowed, essentially saturating the inequality in \cref{eq:minLambda}.\footnote{Technically we assume $\Lambda$ to take the minimum value of the function on the right hand side of \eqref{eq:minLambda} throughout the inflationary trajectory, although this distinction will not be important.} This allows us to obtain a conservative estimate of the swampland bound; $\Lambda$ could certainly be larger, in which case the de Sitter conjecture \eqref{eq:dSconj} is more easily satisfied.

Let us phrase the de Sitter conjecture \eqref{eq:dSconj} as $R\geq c$ for some $c\sim\mathcal{O}(1)$, where the ratio $R$ is
\begin{equation}
R\equiv \frac{1}{\Mp^2}\frac{\mathcal{G}^{ab}\partial_a V\partial_bV}{V^2}.
\end{equation}
Evaluating this for the two-field extension of $P(\phi,X)$ we have
\begin{equation}
R = \Mp^2\frac{\frac{(P_\phi-\chi P_{\phi\chi})^2}{P_\chi}+\Lambda^6\chi^2 P_{\chi\chi}^2}{(P-\chi P_\chi)^2}.
\end{equation}
We assume that on-shell $\chi\approx X$, so that we can evaluate the right-hand side in terms of $\phi$ and $X$ rather than $\phi$ and $\chi$, and write this along the solution of interest---such as an inflationary trajectory---as
\begin{equation}
R \approx \Mp^2\frac{\frac{(P_\phi-X P_{\phi X})^2}{P_X}+X P_{XX}\Box X}{(P-X P_X)^2}, \label{eq:R}
\end{equation}
where, as discussed above, we have taken $\Lambda$ to have its minimum allowed size.

\subsection{Swampland criterion during inflation}

We now proceed to evaluate the expression \eqref{eq:R} on inflationary trajectories. The aim is to derive and collect useful expressions in terms of slow-roll parameters and model parameters that will reduce \cref{eq:R} to the much simpler form \eqref{eq:swampland-PX}. We assume a spatially homogeneous and isotropic solution, and a spatially flat metric,
\begin{equation}
\phi=\phi(t),\qquad \dd s^2 = -\dd t^2+a^2(t)\dd\vec{x}^2.
\end{equation}
The Friedmann equations are
\begin{align}
3\Mp^2H^2 &= -P+2XP_X, \\
\Mp^2\dot H  &= -XP_X,
\end{align}
where overdots denote time derivatives, $H=\dd \ln a/\dd t$, and $X=\dot\phi^2/2$, while the scalar equation of motion is
\begin{align}
P_\phi &= a^{-3}\frac{\dd}{\dd t}\left(a^3P_X\dot\phi\right) \nonumber\\
&= \ddot\phi\left(P_X+\dot\phi^2P_{XX}\right)+\dot\phi^2P_{\phi X}+3HP_X\dot\phi. \label{eq:scalareom}
\end{align}
The Friedmann equations can be combined to obtain an exact expression for the Hubble slow-roll parameter,
\begin{equation}
\varepsilon\equiv-\frac{\dot H}{H^2} = \frac{XP_X}{\Mp^2H^2}.\label{eq:SR-epsilon}
\end{equation}
We demand a quasi-de Sitter phase, so this quantity should be small, $\varepsilon\ll1$.\footnote{Its magnitude is model-dependent. For instance, $\varepsilon\sim n_s-1$ in standard slow-roll inflation, with $n_s$ the tilt of the scalar power spectrum, while in DBI inflation, a prominent example of inflation driven by non-trivial kinetic terms of a $P(\phi,X)$ form, we have instead $\varepsilon\sim\sqrt{n_s-1}$ \cite{Silverstein:2003hf,Alishahiha:2004eh}.} As usual we also insist that its first derivative be small enough to allow inflation to persist for a sufficient number of $e$-foldings, i.e., $\eta\ll1$ with
\begin{equation}
\eta\equiv \frac{\dot\varepsilon}{H\varepsilon}.
\end{equation}
Taking a derivative of \cref{eq:SR-epsilon} and using the definition of $\eta$, we find an expression which will be useful later,
\begin{equation}
\eta-2\varepsilon=\frac{\ddot\phi}{H}\left(\frac2{\dot\phi}+\frac{P_{XX}}{P_X}\dot\phi\right)+\frac{P_{\phi X}}{P_X}\frac{\dot\phi}H. \label{eq:SR-scalar}
\end{equation}
We will also find it convenient to replace $P_X$ in the Friedmann equation using $\varepsilon$, obtaining
\begin{equation}
(3-2\varepsilon)H^2=-\frac{P}{\Mp^2}. \label{eq:SR-fried}
\end{equation}
Note that all of these equations are exact: we have yet to use the slow-roll approximation.

In addition to the slow-roll parameters, which measure the deviation from exact de Sitter, we will introduce a pair of parameters to quantify the deviation from the canonical scalar, for which $P(\phi,X)=X-V(\phi)$. We choose convenient dimensionless parametrizations of $P_{XX}$ and $P_{\phi X}$,
\begin{align}
c_s^2 &\equiv \frac{P_X}{P_X+2XP_{XX}},\\
\beta &\equiv \frac{P_{\phi X}}{P_X}\frac{\dot\phi}{H}.
\end{align}
The first of these, $c_s^2=P_X/\rho_X$, is the usual sound speed, while $\beta$ is chosen to simplify expressions involving $P_{\phi X}$. For a canonical kinetic term we have $\beta=0$ and $c_s=1$, and deviations from these values signal that the non-canonical kinetic structure is important. It is worth mentioning here that, as we pointed out earlier, technically, our UV-extension of the original theory is only valid as long as $P_{\chi\chi}\neq0$ along the trajectory of interest. To leading order in the EFT this condition reads $P_{XX}\neq0$, which we can see is satisfied as long as $P_X\neq0$ (which in turn implies $\varepsilon\neq0$), $X$ is finite, and $c_s<1$. All of these conditions are satisfied by the models under consideration.\footnote{Note also that, even if this were not the case, the fact that $P_{\chi\chi}$ differs from $P_{XX}$ by some higher-order terms in the EFT indicates that the analysis would probably be valid even if the trajectory did encounter a point with $P_{XX}=0$.}

These parameters allow us to write the scalar equation of motion \eqref{eq:scalareom} in a simple form reminiscent of the usual canonical expression,
\begin{equation}
P_X\left[c_s^{-2}\ddot\phi+(3+\beta)H\dot\phi\right]=P_\phi. \label{eq:scalareom2}
\end{equation}
In canonical slow-roll inflation we usually drop the first term, as $\ddot\phi\ll H\dot\phi$. This is a consequence of \cref{eq:SR-scalar}, which for $P(\phi,X)=X-V(\phi)$ becomes $\ddot\phi=(\eta/2-\varepsilon)H\dot\phi$. It is important to understand the conditions under which this continues to hold in the more general $P(\phi,X)$ setting. Let us rewrite \cref{eq:SR-scalar} in terms of $c_s^2$ and $\beta$,
\begin{equation}
\frac{\ddot\phi}{H\dot\phi} = \frac{\eta-2\varepsilon-\beta}{1+c_s^{-2}}, \label{eq:ddotphi}
\end{equation}
where, as already mentioned, we still have not used any slow-roll approximation. Note that, for $0<c_s^2\leq1$, the factor $1/(1+c_s^{-2})$ ranges monotonically from 0 to 1/2, and is approximately $c_s^2$ for $c_s^2\ll1$. This factor is therefore $\mathcal{O}(c_s^2)$ for all $c_s$. We see that $\ddot\phi\ll H\dot\phi$ still holds during slow roll as long as $\beta c_s^2$ is small.\footnote{In DBI inflation, which we discuss in the next section, this quantity is $\mathcal{O}(\varepsilon)$.}

Finally we address the factor $\Box X = \ddot X + 3H\dot X$ appearing in the expression \eqref{eq:R} for $R$. We will assume that $\ddot X\ll H\dot X$. This follows from \cref{eq:ddotphi} during slow-roll if the factor $\beta(1+c_s^{-2})^{-1}\sim\mathcal{O}(\beta c_s^2)$ is small (so that $\ddot\phi\ll H\dot\phi$) and roughly constant over a Hubble time.\footnote{To see this, define $\alpha\equiv\beta(1+c_s^{-2})^{-1}$. We then have $\dot X=\dot\phi\ddot\phi=2\alpha HX$, so $\ddot X\ll H\dot X$ as long as $\alpha\ll1$ and $\dot\alpha\ll H\alpha$.} Under these assumptions, during slow roll we find
\begin{equation}
\Box X \approx -3H\dot X=-3H\dot\phi\ddot\phi.
\end{equation}

Using the expressions presented in this section, we can simplify \cref{eq:R} to find, at leading order in slow-roll,
\begin{equation}
\boxed{R\approx \varepsilon\left(2-\frac{\mb}3+\frac{\mb^2}{18}\right),} \label{eq:swampland-PX}
\end{equation}
where we have defined
\begin{equation}
\mb\equiv\frac{1-c_s^2}{1+c_s^2}\beta.
\end{equation}
Note that $R$ depends on $c_s^2$ and $\beta$ only through this particular combination. We emphasize that the only assumptions we have made in deriving \cref{eq:swampland-PX} are Hubble slow-roll ($\varepsilon,\eta\ll1$) and slow scalar evolution ($\ddot\phi\ll H\dot\phi$).\footnote{Since we are keeping $\beta$ and $c_s$ general, we may be keeping terms in \cref{eq:swampland-PX} which are similar in size to slow-roll terms that we have dropped. The point is to remain agnostic about the size of these model parameters while making use of Hubble slow roll, which is required for a quasi-de Sitter phase.}

The first term in \cref{eq:swampland-PX} is $\mathcal{O}(\epsilon)$, so the only way to satisfy the swampland bound---that is, to have $R>\mathcal{O}(1)$---is to increase the value of $\mb$,
\begin{equation}
\mb>\mathcal{O}\left(\frac1{\sqrt{\varepsilon}}\right)\gg1,
\end{equation}
so that the third (and possibly second) term is large. Our goal therefore becomes to understand the circumstances under which this condition is satisfied. Because we have assumed $c_s^2\beta\ll1$, which is required in order that $\dot\phi$ evolve slowly over a Hubble time, we can only have $\mb\gg1$ if the sound speed is small, $c_s^2<\mathcal{O}(\sqrt\varepsilon)\ll1$. Then $\mb\approx\beta$, so the swampland criterion requires
\begin{equation}
\beta\gtrsim\frac{1}{\sqrt\varepsilon}.
\end{equation}

In the next section we will consider DBI inflation. We will see that on inflationary solutions the sound speed is small, $c_s^2\ll1$, while $\beta$ is large, with the combination $c_s^2\beta\sim\mathcal{O}(\varepsilon)$, implying
\begin{equation}
R\sim\mathcal{O}\left(\frac{\varepsilon^3}{c_s^4}\right).
\end{equation}
We will find, therefore, that the swampland criterion is satisfied, $R\gtrsim\mathcal{O}(1)$, if $c_s$ is sufficiently small to overcome the slow-roll suppression in the numerator.

\section{DBI}
\label{sec:DBI}

In this section we apply the de Sitter conjecture to Dirac-Born-Infeld (DBI) inflation, a particularly prominent, physically-motivated example of inflation with a $P(\phi,X)$ action.\footnote{For earlier work applying the swampland conjectures to DBI inflation, see \rcite{Seo:2018abc,Mizuno:2019pcm}.} One attraction of DBI inflation is that it can be realized as a brane inflation scenario in type IIB string theory, with a D3-brane traveling down a warped throat at relativistic speeds \cite{Silverstein:2003hf,Alishahiha:2004eh}. Besides its motivation in string theory, this model is also appealing because it is a sensible EFT from the bottom-up perspective, since its structure is protected from radiative corrections by a nonlinear symmetry \cite{deRham:2010eu,Goon:2011qf,Babic:2019ify}. Note that the DBI model is constrained by observations, predicting for instance large non-Gaussianities when the sound speed is small \cite{Chen:2006nt}; our focus here is on a proof of concept rather than advocating for a specific inflationary model. As in the previous section we will focus on inflationary trajectories.

The action for DBI inflation is given by \cref{eq:action} with
\begin{equation}
P(\phi,X) = -\frac{1}{g_s}\left(\frac{1}{f(\phi)}\sqrt{1-2f(\phi)X}+V(\phi)\right).
\end{equation}
Here $f(\phi)$ is the (squared) warp factor of the throat, and $V(\phi)$ is the potential. The form of $f(\phi)$ depends on the geometry of the throat; for example for a pure AdS$_5$ throat of radius $R$, we have $f(\phi)=\lambda/\phi^4$ with $\lambda\equiv R^4/\alpha'^2$ \cite{Silverstein:2003hf,Alishahiha:2004eh}. We will leave both the warp factor and the potential general, and return to this specific example at the end.

The distinguishing feature of DBI inflation is ``D-cceleration," an alternative mechanism to potential slow roll. The brane is taken to be moving near the bulk speed of light, imposing a speed limit that leads to slow roll, even if the potential is too steep to allow slow roll in the presence of a canonical kinetic term. From the perspective of the 4D theory this speed limit is imposed by the reality of the square-root term in the action, which requires $2f(\phi)X<1$.

It is convenient to introduce the 5D Lorentz factor $\gamma$,
\begin{equation}
\gamma\equiv\frac{1}{\sqrt{1-2fX}},
\end{equation}
in terms of which the condition for D-cceleration is $\gamma\gg1$. This implies that, to leading order in slow roll,
\begin{equation}
X\approx\frac{1}{2f(\phi)}.
\end{equation}
Calculating the sound speed we find
\begin{equation}
c_s^2 = \frac{1}{\gamma^2}\ll1,
\end{equation}
so that $\mb\approx\beta$, which in turn is
\begin{align}
\beta &= \gamma^2Xf'\frac{\dot\phi}{H} \nonumber\\
&\approx \frac12\frac{\gamma^2}{H}\frac{f'}{f^{3/2}},
\end{align}
where in the second line we have used $X\approx1/(2f)$. The slow-roll parameter $\varepsilon$ is
\begin{align}
\varepsilon &= \frac{XP_X}{\Mp^2H^2} \nonumber\\
&\approx \frac{1}{\Mp^2H^2}\frac\gamma\gs\frac{1}{2f},
\end{align}
where we have used $P_X=\gamma/\gs$, which holds exactly.

These expressions can be written more clearly in terms of the free parameters and functions of the model. We can eliminate $H$ using the Friedmann equation: the potential energy dominates, so we have
\begin{equation}
H^2 \approx \frac{1}{3\gs \Mp^2}V.
\end{equation}
We can also write $\gamma$ in terms of the potential and warp factor. We write the acceleration equation as
\begin{align}
\dot H &= -\frac{1}{2\gs \Mp^2f}\left(\gamma-\frac1\gamma\right)\nonumber \\
&\approx -\frac{\gamma}{2\gs \Mp^2f},
\end{align}
where in the second line we have used $\gamma\gg1$. Taking a time derivative of the Friedmann equation we also find
\begin{equation}
\dot H \approx \frac{V'}{2\sqrt{3\gs}\Mp\sqrt{Vf}}.
\end{equation}
Comparing these, we then obtain an expression for $\gamma$,
\begin{equation}
\gamma\approx\sqrt{\frac\gs3}\Mp\sqrt f\frac{|V'|}{\sqrt V}. \label{eq:gamma}
\end{equation}

Note that we can now calculate $c_s^2\beta$ (which we have seen needs to be small in order to have $\ddot\phi\ll H\dot\phi$ and $\ddot X\ll H\dot X$), finding
\begin{equation}
c_s^2\beta \sim \varepsilon \frac{f'}{f}\frac{V'}{V}.
\end{equation}
In regions of field space where $f(\phi)$ and $V(\phi)$ are, or can be approximated by, power laws with order-unity exponents, then we find $c_s^2\beta\sim\mathcal{O}(\varepsilon)\ll1$.

Recall (cf. \cref{eq:swampland-PX}) that the swampland criterion is given by $R>\mathcal{O}(1)$, with
\begin{equation}
R\approx \varepsilon\left(2-\frac{\beta}3+\frac{\beta^2}{18}\right),
\end{equation}
where we have used the fact that $c_s^2\ll1$ for D-ccelerating DBI theories implies that $\mb\approx\beta$. The first term is clearly small during slow roll. The other two terms are, to leading order in slow roll,
\begin{align}
\varepsilon\beta &\approx\frac14\gs^2\Mp^4\frac{f'}{f}\left|\frac{V'}{V}\right|^3,\\
\varepsilon\beta^2 &\approx\frac{1}{16}\frac{\gs^4\Mp^8}{\varepsilon}\left(\frac{f'}{f}\right)^2\left(\frac{V'}{V}\right)^6 \ .
\end{align}
If the swampland criterion is satisfied, then the term proportional to $\varepsilon\beta^2$ must be dominant. If it were not, i.e., if $\varepsilon\beta^2<\varepsilon\beta$, then $\beta$ would be small, implying $R\approx2\varepsilon\ll1$. In this r\'egime we therefore have
\begin{equation}
R \approx \frac{1}{18}\varepsilon\beta^2 \approx \frac{1}{288}\frac{\gs^4\Mp^8}{\varepsilon}\left(\frac{f'}{f}\right)^2\left(\frac{V'}{V}\right)^6.
\end{equation}

Of course, whether or not this is larger than unity is model-dependent. We can rephrase this result in two more physically-illuminating ways. One is to consider power-law forms for the warp factor and potential,
\begin{equation}
f(\phi) = \lambda \phi^p,\qquad V(\phi) = \mu \phi^q,
\end{equation}
in which case we have
\begin{equation}
R \approx \frac{p^2q^6}{288}\frac{\gs^4}{\varepsilon}\left(\frac\Mp\phi\right)^8,
\end{equation}
or, in terms of the canonically-normalized field $\phi_\mathrm{c}\equiv\phi/\sqrt{\gs}$,
\begin{equation}
R \approx \frac{p^2q^6}{288}\frac1\varepsilon\left(\frac{\Mp}{\phi_\mathrm{c}}\right)^8,
\end{equation}
The simplest example falls into this category: a pure AdS throat with $f(\phi)=\lambda/\phi^4$ and a standard mass term $V(\phi)=m^2\phi^2$, in which case the prefactor $p^2q^6/288$ becomes order unity,
\begin{equation}
R \approx \frac{32}{9}\frac{1}{\varepsilon}\left(\frac{\Mp}{\phi_\mathrm{c}}\right)^8.
\end{equation}
We see that when the warp factor and potential are power laws, the swampland criterion is satisfied when the canonically-normalized field takes on sub-Planckian values.\footnote{The suppression by $1/\varepsilon$ does not significantly affect this conclusion: for $\varepsilon$ in the range of $10^{-1}$--$10^{-2}$, $\varepsilon^{1/8}$ is close to unity. Similarly, it is difficult for the prefactor $p^2q^6/288$ to affect this bottom line; note that $288^{1/8}\approx2.03$. On the other hand, if $\phi_c$ is even double the Planck mass, then $R$ is suppressed by a factor of $2^8$.}

Another useful approach is to compare the swampland criterion to standard slow-roll inflation with the same potential, for which
\begin{equation}
R_\mathrm{SR} = \Mp^2\left(\frac{V'}{V}\right)^2.
\end{equation}
In the D-cceleration r\'egime of DBI, $R$ is enhanced by a factor of
\begin{equation}
\frac{R}{R_\mathrm{SR}} = \frac{\Mp^6\gs^4}{288}\frac1\varepsilon\left(\frac{f'}{f}\right)^2\left(\frac{V'}{V}\right)^4.
\end{equation}
We see that the non-canonical kinetic structure, appearing through $f(\phi)$, is crucial for evading the swampland bound, since it is large gradients in $f(\phi)$ that enhance $R$. We emphasize that one should be careful when comparing $V'/V$ to its single-field slow-roll counterpart, as the D-cceleration mechanism allows for inflation with rather larger values of the inflaton mass (or, more generally, with steeper potentials) than does standard single-field slow roll \cite{Alishahiha:2004eh}.

\section{Turning estimates}
\label{sec:turning}

We have seen that $P(\phi,X)$ theories can have quasi-de Sitter phases which evade the swampland conjecture \eqref{eq:dSconj}. The analysis requires these theories to be rephrased in a multi-field form via the introduction of an auxiliary field which is given (small) dynamics. In general, multi-field inflation theories are able to evade the swampland bound with what is known as large {\it turning} in field space \cite{Achucarro:2018vey}. In contrast to single-field inflation, in a multi-field context the fields can take a variety of trajectories through field space, rather than being required to follow gradients of the potential; trajectories develop angular momentum in field space, or turning, when the field trajectory is misaligned with the gradient flow of the potential. This severs the direct link between the Hubble and potential slow-roll parameters, allowing the field value to stay approximately constant even on a steep potential, driving acceleration. In this section we calculate the amount of turning on inflationary trajectories in the multi-field picture in $P(\phi,X)$ inflation, and show that, in agreement with \rcite{Achucarro:2018vey}, the evasion of the swampland bound is due to large turning in field space in the two-field UV extension.

Our analysis will follow the geometric approach summarized in \rcite{Achucarro:2018vey}. Let us package the fields into a field-space vector $\Phi^a=(\phi,\chi)$.\footnote{Technically the units are mixed, since $\chi$ has units of mass$^4$, but our calculations will be self-consistent, since the field-space metric has correspondingly mixed units.} We define the norm of the time derivative of $\Phi^a$ by
\begin{equation}
\dot\Phi \equiv \sqrt{\mathcal{G}_{ab}\dot\Phi^a\dot\Phi^b} = \sqrt{P_X\dot\phi^2+\frac{1}{\Lambda^6}\dot X^2}. \label{eq:Phidot}
\end{equation}
As in the previous sections, we assume we are in the low-energy r\'egime of the EFT and freely use $\chi\approx X$. Next we define a covariant time derivative $D_t$ by
\begin{equation}
D_t V^a \equiv \dot V^a + \Gamma^a_{bc}V^b\dot\Phi^c,
\end{equation}
where $V^a$ is any vector in field space and $\Gamma^a_{bc}$ are the Christoffel symbols associated with the field-space metric, whose only nonzero components are
\begin{equation}
\Gamma^\phi_{\phi\phi} = \frac12\frac{P_{\phi X}}{P_X},\qquad \Gamma^\phi_{\chi\phi} = \frac12\frac{P_{XX}}{P_X},\qquad \Gamma^\chi_{\phi\phi} = -\frac12\Lambda^6P_{XX}.
\end{equation}
Defining the unit tangent vector,
\begin{equation}
T^a \equiv \frac{\dot\Phi^a}{\dot\Phi},
\end{equation}
we can finally define the angular velocity in field space,
\begin{align}
\Omega &\equiv |D_tT| \nonumber\\
&= \sqrt{P_X (D_tT^\phi)^2 + \frac{1}{\Lambda^6}(D_tT^\chi)^2},
\end{align}
where the components of $D_tT^a$ are
\begin{align}
D_tT^\phi &= \left(\frac{\dot\phi}{\dot\Phi}\right)\dot{\bigg.} + \frac{1}{2P_X}\frac{\dot\phi}{\dot\Phi}\left(P_{\phi X}\dot\phi+2P_{XX}\dot X\right),\\
D_tT^\chi &= \left(\frac{\dot X}{\dot\Phi}\right)\dot{\bigg.} - \frac12\Lambda^6P_{XX}\frac{\dot\phi^2}{\dot\Phi}.
\end{align}

We begin by estimating $\dot\Phi$. Per \cref{eq:Phidot}, this has two terms, one from the $\phi$ sector and one from the $\chi$ sector. Using the expressions and approximations presented in \cref{sec:swamp} we find, to leading order in slow roll,
\begin{equation}
\frac{\Lambda^{-6}\dot X^2}{P_X\dot\phi^2} = \frac\mb6.
\end{equation}
If the swampland criterion is satisfied then we have $\mb>\mathcal{O}(\varepsilon^{-1/2})\gg1$, in which case we find
\begin{equation}
\dot\Phi\approx\frac{\dot X}{\Lambda^3}.
\end{equation}
We can perform a similar analysis for the turning rate, $\Omega=|D_tT|$. We start with the term $\dot\phi/\dot\Phi$ which appears in $D_tT^\phi$, and which in slow roll we can straightforwardly calculate as,
\begin{equation}
\left(\frac{\dot\phi}{\dot\Phi}\right)^2 \approx\frac{6}{\mb P_X}.
\end{equation}
Assuming $\mb$ is approximately constant, we find
\begin{align}
\left(\frac{\dot\phi}{\dot\Phi}\right)\dot{\bigg.} &\approx -2\frac{\dot\phi}{\dot\Phi}\frac{\dot{P}_X}{P_X}\nonumber\\
&\approx-\frac{\beta}{1+c_s^{-2}}H\frac{\dot\phi}{\dot\Phi}.
\end{align}
Recall from \cref{sec:swamp} that the same combination of $\beta$ and $c_s^2$ appears in the ratio $\ddot\phi/(H\dot\phi)$, and so under the same assumptions we have been making, we see that $\dot\phi/\dot\Phi$ varies slowly over a Hubble time. Using this in our expression for $D_tT^\phi$ we see that the $(\dot\phi/\dot\Phi)\dot{}$ term is subdominant.\footnote{In particular, it is suppressed by a factor of $c_s^2$, which we have seen is necessarily small if $R>1$ and $\ddot\phi\ll H\dot\phi$.} The time-derivative term $D_tT^\chi$ is also negligible: $\dot X/\dot\Phi\approx \Lambda^3$, which is constant by definition. With these approximations we can then calculate the ratio of terms appearing in $\Omega$,
\begin{align}
\frac{\Lambda^{-6}(D_tT^\chi)^2}{P_X (D_tT^\phi)^2} &\approx \frac{6}{\beta}\frac{1-c_s^4}{(1-3c_s^2)^2} \nonumber\\
&= \frac6\mb\left(\frac{1-c_s^2}{1-3c_s^2}\right)^2.
\end{align}
If the swampland criterion is satisfied then this is much smaller than unity, and so we find
\begin{equation}
\Omega^2 \approx P_X(D_t T^\phi)^2.
\end{equation}
Note that the $\chi$ contribution dominates $\dot\Phi$, while the $\phi$ contribution dominates $\Omega$.

To evaluate the magnitude of this we should compare it to the Hubble rate, which has the same dimensions,
\begin{align}
\frac{\Omega^2}{H^2} &\approx \frac32\beta\frac{(3c_s^2-1)^2}{1-c_s^4}\nonumber\\
&= \frac32\mb\left(\frac{1-3c_s^2}{1-c_s^2}\right)^2.
\end{align}
If $R>\mathcal{O}(1)$ then $c_s^2\ll1$ and $\beta\gg1$, so the turning is large,
\begin{equation}
\frac{\Omega^2}{H^2}\sim\mathcal{O}(\beta)\gg1.
\end{equation}
We conclude that, from the perspective of the two-field picture, DBI inflation has large turning when the de Sitter swampland criterion is satisfied.

These considerations show that $P(\phi,X)$ theories fit into the analysis of \rcite{Achucarro:2018vey}, which studied, in greater generality, multi-field inflation and the swampland conjectures in the presence of large $\Omega$, taking into account both the de Sitter conjecture \eqref{eq:dSconj} and the distance conjecture \eqref{eq:distconj}. They find that satisfying both of these swampland conjectures simultaneously requires
\begin{equation}
\Omega\gtrsim180H.
\end{equation}
In the context of $P(\phi,X)$ inflation this imposes
\begin{equation}
\sqrt\beta\gtrsim150.
\end{equation}

\section{Conclusions}
\label{sec:conc}

We have applied the de Sitter swampland conjecture of \rcite{Obied:2018sgi} to scalar field theories with non-canonical kinetic structures of the $P(\phi,X)$ type. This conjecture is phrased in terms of non-linear sigma models whose kinetic terms are quadratic in derivatives. We fit $P(\phi,X)$ theories into this framework by introducing an auxiliary field and giving it a small kinetic term, so that solutions to the $P(\phi,X)$ theory remain approximate solutions to the two-field theory. This can be seen as embedding the $P(\phi,X)$ theory in a multi-field UV extension.

The de Sitter conjecture translates into a bound on the function $P(\phi,X)$ and its derivatives. We have analyzed this condition on quasi-de Sitter inflationary solutions in conjunction with the Hubble slow-roll approximation,\footnote{Note that the bound should hold over all field space, not just on the inflationary trajectory.} and have applied them to a prominent string-inspired inflationary model of the $P(\phi,X)$ form: DBI inflation. We find that DBI inflation can satisfy the swampland bound given potentials which would violate the bound in the presence of a canonical kinetic structure. We have shown that when these models satisfy the swampland conjecture, they do so because of \emph{turning} in field space, in agreement with other examples of multi-field inflation which do not violate the swampland bound \cite{Achucarro:2018vey}.

While the techniques we have used in this paper have proven extremely useful for extending swampland bounds to one set of theories with nontrivial kinetic terms, they are not applicable, at least in a simple way, to all such theories. In particular, it would be very interesting to study whether there are methods to extend the same reasoning to theories such as the galileon or Horndeski models, in which higher-derivative terms enter explicitly into the Lagrangians. Thus far we have not uncovered such techniques.

\begin{acknowledgments}
We thank Mark Hertzberg and particularly McCullen Sandora for useful discussions. A.R.S. is supported by DOE HEP grants DOE DE-FG02-04ER41338 and FG02-06ER41449 and by the McWilliams Center for Cosmology, Carnegie Mellon University. The work of M.T. is supported in part by US Department of Energy (HEP) Award DE-SC0013528, and by the Simons Foundation, grant number 658904.
\end{acknowledgments}

\bibliography{refs}

\end{document}